\documentclass[12pt]{iopart}
\usepackage{psfig}

\newcommand\aplt{\ {\raise-.5ex\hbox{$\buildrel<\over\sim$}}\ }
\newcommand\apgt{\ {\raise-.5ex\hbox{$\buildrel>\over\sim$}}\ }
\newcommand\msun{\ensuremath{M_\odot}}
\newcommand\lsun{\ensuremath{L_\odot}}

\begin{document}

\title{Gravitational Dynamics of Large Stellar Systems}

\author{Stephen L. W. McMillan}

\address{Department of Physics, Drexel University, Philadelphia, PA
  19104, USA}

\ead{steve@physics.drexel.edu}

\begin{abstract}
Internal dynamical evolution can drive stellar systems into states of
high central density.  For many star clusters and galactic nuclei, the
time scale on which this occurs is significantly less than the age of
the universe.  As a result, such systems are expected to be sites of
frequent interactions among stars, binary systems, and stellar
remnants, making them efficient factories for the production of
compact binaries, intermediate-mass black holes, and other interesting
and eminently observable astrophysical exotica.  We describe some
elements of the competition among stellar dynamics, stellar evolution,
and other mechanisms to control the dynamics of stellar systems, and
discuss briefly the techniques by which these systems are modeled and
studied.  Particular emphasis is placed on pathways leading to massive
black holes in present-day globular clusters and other potentially
detectable sources of gravitational radiation.
\end{abstract}

\maketitle


\section{Dynamical Evolution in a Nutshell}

\subsection{Evolutionary Time Scales}

The dynamics of a self-gravitating system is defined by two
fundamental time scales.  The {\em dynamical time}, $t_d$, is the
time required for a typical star to cross the system.  It is also the
time scale on which the system establishes virial equilibrium,
$2T+U=0$, where $T$ and $U$ are the kinetic and gravitational
potential energies of the system, respectively.  A convenient formal
definition in terms of conserved quantities is
\begin{equation}
	t_d = \frac{GM^{5/2}}{(-4E)^{3/2}}\,,
\end{equation}
where $M$ is the total mass of the system and $E=T+U$ is the total
energy.  In virial equilibrium, this expression assumes the more
familiar form
\begin{equation}
	t_d = \left(\frac{GM}{R^3}\right)^{-1/2}
	    = 4.7\times10^5\,{\rm yr}
	      \left(\frac{M}{10^3\,\msun}\right)^{-1/2}
	      \left(\frac{R}{1\,{\rm pc}}\right)^{3/2}
\label{eq:dynamical}
\end{equation}
where $R=-GM^2/2U$ is the virial radius and we have added for
reference some astrophysically relevant scales.  In this review we
confine our attention to systems in virial equilibrium (neglecting
such interesting cases as merging star clusters or colliding
galaxies), and take Eq.~(\ref{eq:dynamical}) as our working definition
of the dynamical time scale.

The second fundamental time scale is the {\rm relaxation time}, $t_r$,
on which two-body encounters transfer energy between individual stars
and cause the system to approach thermal equilibrium.  Spitzer (1987)
derives the local expression
\begin{equation}
	t_r = \frac{0.03\langle v^2\rangle^{3/2}}
		   {G^2\langle m\rangle\rho\log\Lambda} \,,
\end{equation}
where $\langle v^2\rangle$ is the velocity dispersion, $\langle
m\rangle$ is the mean stellar mass, $\rho$ is the density,
$N=M/\langle m\rangle$, and $\Lambda\sim0.4N$ for a system in virial
equilibrium.  Replacing all quantities by their cluster-wide
(``half-mass'') averages, we obtain the half-mass relaxation time
\begin{equation}
	t_{rh} = \frac{0.08M^{1/2}R^{3/2}}{G^{1/2}\langle m\rangle\log\Lambda}
	       = \frac{N}{12\log\Lambda}\,t_d\,.
\label{eq:trtd}
\end{equation}

A natural distinction can be drawn between ``collisionless'' systems,
whose long relaxation times mean that they will not undergo
significant internal dynamical evolution during the age of the
universe, and ``collisional'' systems, which evolve significantly in
less than a Hubble time.  Galaxies fall into the former category---a
typical relaxation time in a modest galactic spheroid, with mass
$10^{11}\msun$ and radius $10$ kpc is $\sim3\times10^{16}$ yr.  Star
clusters and some galactic nuclei fall into the latter category---a
globular cluster, for example, with $M\sim10^6\msun$ and $R\sim10$ pc
has $t_r\sim10^{10}$ yr.  The median relaxation time of the Galactic
globular cluster system is $\sim1.2$ Gyr (Harris 1996).  We further
limit our discussion here to collisional systems, in which substantial
internal evolution is expected on time scales less than the age of the
universe.

\subsection{Evolution of Collisional Stellar Systems}

\begin{figure}
  \vskip 0.2in
  \centerline{\psfig{figure=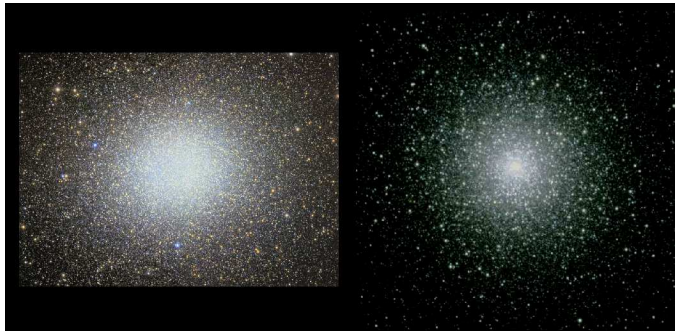,width=3.5in}}
  \caption{Pre-core-collapse ($\omega$ Centauri, left) and
    post-collapse (M15, right) globular clusters (not shown to scale).}
  \label{fig:cluster}
\end{figure}

Self-gravitating systems are inherently unstable, and no final
equilibrium state exists for a star cluster.  The escape of
high-velocity stars (evaporation) and the internal effects of two-body
relaxation, which transfers energy from the inner to the outer regions
of the cluster, precipitate the phenomenon known as {\em core
  collapse} (Antonov 1962, Cohn 1980).  During this phase, the central
portions of the cluster accelerate toward infinite density while the
outer regions expand.  The process is readily understood by
recognizing that, according to the virial theorem, a self-gravitating
system has negative specific heat---reducing its energy causes it to
heat up.  Hence, as relaxation transports energy from the
(dynamically) warmer central core to the cooler outer regions, the
core contracts and heats up as it loses energy.  The time scale for
the process to go to completion (i.e. a core of zero size and formally
infinite density) is $t_{cc}\sim 15 t_{rh}$ for a system of identical
masses, and significantly less in the case of a broad spectrum of
masses (Inagaki 1985; Portegies Zwart and McMillan 2002).  Figure
\ref{fig:cluster} compares the appearance of pre- and post-collapse
systems.

In systems with a mass spectrum, two-body interactions accelerate the
dynamical evolution by driving the system toward energy equipartition,
in which the velocity dispersions of stars of different masses would
have $m\langle v^2\rangle\sim {\rm constant}$.  The result is {\em
  mass segregation}, where more massive stars slow down and sink
toward the center of the cluster on a time scale (Spitzer 1969)
\begin{equation}
	t_s \sim \frac{\langle m\rangle}{m}\,t_{rh} \,.
\label{eq:ts}
\end{equation}

Thus, a collisional stellar system inevitably evolves toward a state
in which the most massive objects become concentrated in the
high-density central core, and this process can occur in less than a
Hubble time.  From an astrophysical point of view, these ``most
massive objects'' are the most interesting objects in the
system---black holes, neutron stars, massive white dwarfs, and binary
star systems.  Dynamical evolution provides a natural and effective
mechanism for concentrating astrophysically interesting objects in
regions of high stellar density, thereby promoting interactions among
them.

These interactions not only create potentially observable objects,
such as X-ray binaries, millisecond pulsars, and (perhaps) sources of
gravitational radiation, but in many cases can also play important
roles in the dynamics of their parent system.  Binary stars, for
example, act as heat sources to the stellar system through
superelastic encounters with other cluster members (Heggie 1975),
ultimately halting and even reversing core collapse (McMillan \etal
1991; McMillan and Hut 1994).  At the same time, they greatly increase
the effective cross sections for close encounters and collisions
between stars and remnants.  The pursuit of a full understanding of
the dynamical evolution of collisional stellar systems therefore
drives us to study not just the gravitational dynamics of the stars
themselves, but also the physics of stellar interactions and
collisions, and the processes driving the stellar and binary evolution
of the resulting objects.


\section{Modeling Dense Stellar Systems}\label{sec:modeling}

The past decade has seen groundbreaking advances in both hardware
design and software development and integration, all of which have
been crucial in advancing our understanding of this field.  Here we
describe a few of the computational techniques currently in use.

\subsection{The State of the Art}

Consistent with our growing understanding of the importance of stellar
and binary interactions in collisional stellar systems, the leading
programs in this area are the various ``kitchen sink'' packages that
combine treatments of dynamics, stellar and binary evolution, and
stellar hydrodynamics within a single simulation.  Of these, the most
widely used are the $N$-body codes {\tt NBODY} (Aarseth 2003) and {\tt
  kira} (e.g. Portegies Zwart \etal 2001), and the Monte-Carlo codes
developed by Giersz (Giersz 1998; see also Giersz and Heggie 2008),
Freitag (Freitag \etal 2006), and by Rasio and coworkers (e.g. Fregeau
\etal 2003).

These codes differ principally in their handling of the large-scale
dynamics, employing conceptually similar approaches to stellar and
binary evolution and collisions.  All use approximate descriptions of
stellar evolution, generally derived from look-up tables based on the
detailed evolutionary models of Eggleton \etal (1989) or Hurley \etal
(2000).  They also rely on semi-analytic or heuristic rule-based
treatments of binary evolution, conceptually similar from code to
code, but significantly different in detail.  In most cases,
collisions are implemented in the simple ``sticky-sphere''
approximation, where stars are taken to collide (and merge) if they
approach within the sum of their effective radii.  The effective radii
may be calibrated using hydrodynamical simulations, and mass loss may
be included in some approximate way.  Freitag's Monte-Carlo code uses
a more sophisticated approach, interpolating encounter outcomes from a
pre-computed grid of SPH simulations (Freitag and Benz 2005).
Small-scale dynamics of multiple stellar encounters, such as binary
and higher-order encounters, may be handled by look-up from
precomputed cross sections or---more commonly---by direct integration,
either in isolation or as part of a larger N-body calculation.  Codes
employing direct integration may also include post-Newtonian terms in
the interactions between compact objects.

$N$-body codes incorporate detailed descriptions of stellar dynamics
at all levels, using direct integration of the individual (Newtonian)
stellar equations of motion for all stars.  Monte-Carlo methods are
designed for efficient computation of relaxation effects in
collisional stellar systems, a task which they accomplish by reducing
stellar orbits to their orbital elements---energy and angular
momentum---effectively orbit averaging the motion of each star.
Relaxation is modeled by randomly selecting pairs of stars and
applying interactions between them in such a way that, on average, the
correct rate is obtained.  This may be implemented in a number of
ways, but interactions are generally realized on time scales
comparable to the orbit-averaged relaxation time.  As a result,
Monte-Carlo schemes can be orders of magnitude faster than direct
$N$-body codes.  Joshi \etal (2000) report an empirical CPU time
scaling of $O(N^{1.4})$ for core-collapse problems (compared with
$N^3$ for $N$-body methods, as discussed below).  To achieve these
speeds, however, the geometry of the system must be simple enough that
the orbital integrals can be computed from a star's instantaneous
energy and angular momentum.  In practice, this limits the approach to
spherically symmetric systems in virial equilibrium, and global
dynamical processes occurring on relaxation (or longer) time scales.

The major attraction of $N$-body methods is that they are
assumption-free, in the sense that all stellar interactions are
automatically included to all orders, without the need for any
simplifying approximations or the inclusion of additional reaction
rates to model particular physical processes of interest.  Thus,
problems inherent to Monte-Carlo methods, related to departures from
virial equilibrium, spherical symmetry, statistical fluctuations, the
form of (and indeed the existence of) phase space distribution
functions, and the possibility of interactions not explicitly coded in
advance, simply do not arise.  The price of this is computational
expense.  Each of the $N$ particles must interact with every other
particle a few hundred times over the course of every orbit, each
interaction requires $O(N)$ force calculations, and a typical
(relaxation time) run spans $O(N)$ orbits (see Eq.~\ref{eq:trtd}).
The resulting $O(N^3)$ scaling of the total CPU time means that, even
with the best time-step algorithms, integrating even a fairly small
system of, say, $N\sim10^5$ stars requires sustained teraflops speeds
for several months (Hut \etal 1988).

The technological solution now in widespread use is the ``GRAPE''
(short for ``GRAvity PipE'') series of machines developed by Sugimoto
and co-workers at Tokyo University (Ebisuzaki \etal 1993).
Abandoning algorithmic sophistication in favor of simplicity and raw
computing power, these machines achieve high speed by mating a
fourth-order Hermite integration scheme (Makino and Aarseth 1992) with
special-purpose hardware in the form of highly parallel, pipelined
``Newtonian force accelerators'' which implement the computation of
all interparticle forces entirely in hardware.  Operationally, the
hardware is simple to program, as it merely replaces the function that
computes the force on a particle by a call to the hardware interface
libraries; the remainder of the user's $N$-body code is unchanged.
The effect of GRAPE on simulations of stellar systems has been nothing
short of revolutionary.  Today, GRAPEs lie at the heart of almost all
detailed $N$-body simulations of star clusters and dense stellar
systems.

Recently, Graphics Processing Units (GPUs) have begun to achieve
speeds and price/performance levels previously attainable only by
GRAPE systems (see Belleman \etal 2008 for a recent GPU implementation
of the GRAPE interface).  It appears that commodity components may be
poised to outpace special-purpose computers in this specialized area
of computational science, just as they have already done in
general-purpose computing.

\subsection{The MODEST Initiative}

The realization that the evolution of dense stellar systems results
not only in interactions among stars, but also among stellar modelers
and their programs, has been a major motivating factor in the
``MODEST'' program ({\tt http://www.manybody.org/modest}).  Short for
MOdeling DEnse STellar systems, MODEST is a loosely knit collection of
groups working on all aspects of the theory and observations of star
clusters, including stellar dynamics, stellar evolution, stellar
hydrodynamics, and cluster formation.  MODEST has hosted some 20
meetings over the past 5 years, providing an invaluable forum for
discussion and collaboration among researchers in this field.

A key goal of MODEST is to provide a software framework for
large-scale simulations of dense stellar systems, within which
existing programs for dynamics, stellar evolution, and hydrodynamics
can be easily coupled.  An important aspect of this is the
incorporation of ``live'' treatments of stellar and binary evolution
and ultimately stellar hydrodynamics directly into kitchen-sink
$N$-body simulations.  Such an undertaking is essential if one wishes
to model the evolution of a dense stellar system, in which stellar
collisions may be commonplace events, creating wholly new channels for
stars to evolve and allowing the formation of stellar species
completely inaccessible by standard stellar and binary evolutionary
pathways.

A pioneering effort to couple stellar dynamics and stellar evolution
was reported by Church (2006), who combined Aarseth's {\tt NBODY6}
with a version of Eggleton's EV (Eggleton 2006).  This heroic
accomplishment hard-coded the two programs into a single application,
providing proof of concept that two such disparate modules could in
fact be successfully merged.  The MODEST approach to code integration,
called ``MUSE,'' for MUltiscale MUltiphysics Scientific Environment
({\tt http://muse.li}), adopts a rather different approach, providing
instead a modular python framework within which programs written by
many different authors, and in many different languages, can
interoperate with as little intrusion as possible into the internal
operation of each (Portegies Zwart \etal 2008).

This model has many advantages.  By providing well defined interfaces
between modules, it allows researchers (or students) to quickly build
real scientific applications using state-of-the-art techniques,
without first having to become experts in the many details of each
module.  Equally important, the MUSE structure allows users to write
generic scripts that do not depend on the details of the algorithms
used, providing, for the first time, ``plug and play'' functionality
that allows different combinations of modules to be combined and
compared.


\section{Black Holes in Star Clusters}

Rather than attempting to describe the many astrophysical consequences
of collisional stellar dynamics, we turn to a specific problem of
particular interest to the relativity community, the possible presence
of black holes in these systems.

Black holes (BHs) are natural products of stellar evolution in massive
stars, and may also result from dynamical interactions in dense
stellar environments.  They can significantly influence the dynamics
of their parent cluster, and may also have important observational
consequences, via their X-ray emission, the production of
gravitational waves, and their effect on the structural properties of
the system in which they reside.  Globular clusters offer particularly
rich environments for the production of black holes in statistically
significant numbers, yet direct evidence for black holes in globulars
is scarce, although several independent lines of investigation now
hint at their presence.  Despite the lack of firm observational
support, the past three decades have seen many theoretical studies of
the formation and dynamics of stellar- and intermediate-mass black
holes in star clusters.


\subsection{Observations of Stellar-Mass Black Holes}

Given the numerous theoretical studies of the formation and dynamical
consequences of BHs in star clusters, and the overwhelming weight of
opinion on the inevitability of BHs as consequences of stellar
evolution, there is remarkably little observational evidence for
stellar-mass (i.e. less than a few tens of solar masses) BHs in
globular clusters.  There appears to be no evidence for such low-mass
BHs in the Galactic globular cluster population, and only one firm
observation of a stellar-mass BH in an extragalactic globular cluster.

The sole extragalactic BH was reported by Maccarone \etal (2007), who
observed a bright X-ray source in a cluster in NGC 4472, a bright
elliptical galaxy in the Virgo cluster.  Its X-ray luminosity of
$\sim4\times10^{39}$ erg/s is the Eddington luminosity for a 35
{\msun} object, and the source shows substantial variability on a time
scale of hours, perhaps indicating a considerably larger mass, so even
this candidate may actually fall into the intermediate-mass black hole
(IMBH) range.  Interestingly, the parent cluster is itself quite
bright ($V=21$, or $L\sim7.5\times10^5\lsun$ at a distance of 16 Mpc)
and lies far (30 kpc) from the center of the host galaxy, perhaps
placing this system in the same category as the leading candidates for
IMBHs in the Milky Way ($\omega$ Centauri: $L\approx
1.1\times10^6\lsun$; Harris 1996) and M31 (G1:
$L\sim2.1\times10^6\lsun$; Meylan \etal 2001), as discussed in
\S\ref{sec:observations}.


\subsection{Intermediate-Mass Black Holes}
\label{sec:observations}

Although there is considerable theoretical uncertainty about how such
objects might form, the observational evidence for IMBHs may actually
be stronger than that for stellar-mass BHs.  Several lines of
reasoning support the assertion that IMBHs exist in globular clusters:
(1) observations of ultraluminous X-ray sources (ULXs), (2) dynamical
modeling, and (3) studies of cluster structure.


\subsubsection{Ultraluminous X-ray Sources}

The bright X-ray source M82 X-1 (Matsumoto and Tsuru 1999; Matsumoto
\etal 2001; Kaaret \etal 2001) is the strongest ULX candidate for an
IMBH.  With a peak luminosity of more than $10^{41}$ erg/s, it is too
bright to be an ordinary X-ray binary, while its location 200 pc from
the center of M82 argues against a supermassive black hole.  This
luminosity is consistent with an accreting compact object of at least
350 solar masses, possibly an IMBH.  The discovery of
$54.4\pm0.9$\,mHz quasi-periodic oscillations (Strohmayer and
Mushotsky 2003) supports this view.

\begin{figure}
  \vskip 0.2in
  \centerline{\psfig{figure=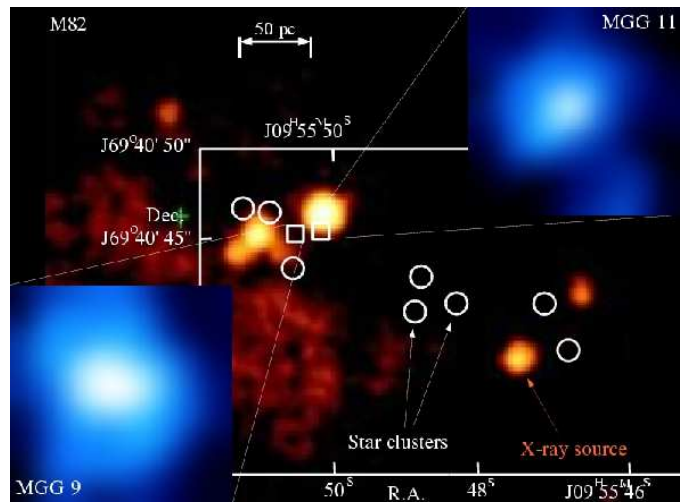,width=3.5in}}
  \caption{Combined HST and Chandra images of the region containing
    M82-X1 and MGG-11.  The X-ray observations of Matsumoto \etal
    (2001) are shown in color; M82 X-1 is near the center of the
    image.  The star clusters from Table 3 of McCrady \etal (2003),
    are indicated by circles.  The positions of the clusters MGG-9 and
    MGG-11 studied by Portegies Zwart \etal 2004 are indicated by
    squares.  Infrared images from the McCrady \etal observations are
    presented in the upper right (MGG-11) and lower left (MGG-9)
    corners.  The quasi-periodic oscillator is not shown because of
    its low (7 arcsecond) positional accuracy, but its position is
    consistent with the X-ray source in MGG-11.  (From Portegies Zwart
    \etal 2004.)}
  \label{fig:M82}
\end{figure}

An intriguing aspect of M82 X-1 is its apparent association with the
young dense cluster MGG-11 (McCrady \etal 2003).  Figure
\ref{fig:M82} (from Portegies Zwart \etal 2004) shows superimposed
near-IR (HST) and X-ray (Chandra) images of the region containing M82
X-1 and MGG-11.  (The offset between the infrared cluster and the
X-ray source is consistent with the absolute pointing accuracies of
the two telescopes.)  The cluster age is between 7 and 12 Myr.
Portegies Zwart \etal (2004) found that such an association would be
consistent with the scenario described in \S\ref{sec:formation} for
runaway stellar growth in a dense cluster.

Numerous authors have noted that high X-ray luminosity is by no means
conclusive evidence of an IMBH (see Miller and Colbert 2004 for a
review of some alternative possibilities).  Soria (2006, 2007) points
out that most ULXs may in fact be consistent with the high-luminosity
tail of the X-ray binary luminosity function, and that ULXs are
generally not associated with young star clusters.  Nevertheless, the
runaway collision scenario has become the ``standard'' mechanism for
IMBH formation in clusters, against which others are assessed.

\subsubsection{Dynamical Modeling of Cluster Velocity Structure}

Currently the most definitive statements about IMBHs in globular
clusters have come from Gebhardt and collaborators (Gerssen \etal
2002, 2003; Noyola \etal 2006; Noyola 2008), based on axisymmetric,
three-integral dynamical models of cluster potentials, using full
line-of-sight velocity information to constrain the model fits.  IMBH
masses have been reported for the Galactic globular clusters M15 and
$\omega$ Centauri (Fig.~\ref{fig:cluster}), and for the cluster G1 in
M31, although these results have not been without controversy.

Gerssen \etal (2002) reported dynamical evidence for a
$4\pm2\times10^3\msun$ IMBH in M15.  This result was criticized by
Baumgardt \etal (2003a), who pointed out that a ``standard''
dynamically evolved cluster model would yield similar results, and was
marred by a crucially mislabeled figure in an earlier study of M15's
dynamics.  Subsequently the estimate of the IMBH mass was reduced to
$2\pm2\times10^3\msun$ (Gerssen \etal 2003).  It now seems clear, from
a variety of different arguments, that the core of this highly
centrally concentrated cluster does contain on the order of
$1000\msun$ of non-luminous matter (Dull \etal 1997, Baumgardt \etal
2003a, Phinney 1993).  However, given the age and likely history of
M15, the dark material most plausibly consists of stellar
remnants---neutron stars and/or heavy white dwarfs---providing a much
more natural explanation of the invisible mass.  The models do not
rule out an IMBH at the $\sim500-1000\msun$ level, but present no
compelling reason to conclude that one exists.

Much firmer evidence for IMBHs is found in G1 and $\omega$ Centauri,
the largest globular clusters in M31 and the Milky Way, respectively.
Both clusters are large enough that their relaxation times are long,
making it unlikely that the dynamics of stellar evolution products
could mimic the effect of a central IMBH, as in M15.

In G1, a cluster with mass $\sim 1.5\times10^7\msun$, situated some 40
kpc from the center of M31 (Meylan \etal 2001), the dynamical models
yield a black hole mass of $1.8\pm0.5\times10^4\msun$ (Gebhardt \etal
2002, 2005).  The initial results were also questioned by Baumgardt
\etal (2003b), in part because the $0.013^{\prime\prime}$-radius
sphere of influence of the supposed IMBH lies well inside the central
pixel of the HST image, but also because plausible $N$-body models
without black holes could reproduce the cluster's observed surface
density and velocity dispersion profiles quite well.  However,
Gebhardt \etal (2005) have argued that the use of the full velocity
distribution, and not just its lowest moments, is essential in order
to properly define the cluster potential, and that the $N$-body
simulations simply lack the resolution necessary for them to be
meaningfully compared with the observational data.

Support for the Gebhardt \etal result has come from the recent
detection of X-ray emission from G1 (Pooley and Rappaport 2006).  The
observed luminosity $L_X\sim2\times10^{36}$ erg/s is consistent with
Bondi-Hoyle accretion of intracluster gas onto an IMBH in the relevant
mass range.  Further support comes from radio observations of G1
(Ulvestad \etal 2007) which imply a radio to X-ray flux ratio
consistent with a $2\times10^4\msun$ black hole (Merloni \etal 2003).

Noyola \etal (2006) have reported an IMBH mass of
$4\pm1\times10^4\msun$ in $\omega$ Centauri, a globular cluster with
mass $\sim5\times10^6\msun$, lying $\sim 6$ kpc from the center of the
Milky Way.  Interestingly, both G1 and $\omega$ Centauri lie near the
low-mass extension of the ``$M-\sigma$'' relation for active galactic
nuclei (Merritt and Ferrarese 2001; Tremaine \etal 2002), reinforcing
the suspicion that G1 and $\omega$ Centauri are in fact not ``real''
globular clusters, but rather are the cores of dwarf spheroidal
systems stripped by the tidal fields of M31 and the Milky Way (Meylan
\etal 2001, Freeman 1993).  This possibility does not explain the
origin of the IMBHs, but it obviously moves the question into a very
different arena.


\subsubsection{Indirect Evidence for Central Black Holes}
\label{sec:indirect}

Noyola and Gebhardt (2006) find that a a surprisingly large fraction
($\sim$25\%) of globular clusters hitherto thought to have
``classical'' cores in fact have shallow power-law surface brightness
profiles in their central regions.  The detection of these weak cusps
is due in large part to high-resolution HST observations of the
innermost arcsecond of these systems.  These observations remain
somewhat controversial, but are in good agreement with theoretical
simulations by Baumgardt \etal (2005) of clusters containing central
black holes comprising $\sim$0.1-1\% of the total cluster mass, and
have been interpreted as indirect indicators of IMBHs in these
clusters.  Ongoing detailed dynamical studies by Noyola \etal (2008)
of selected clusters from their earlier study, including NGC 2808, 47
Tucanae, and the ``weak cusp'' systems M54 and M80 should shed much
further light on the dynamics of these intriguing systems.

Curiously, although dense stellar systems are among the most promising
environments for the formation of IMBHs (see \S\ref{sec:formation}),
they may not be the best place to look for evidence of massive black
holes.  Baumgardt \etal (2005) and Heggie \etal (2007) have pointed
out that core-collapse clusters such as M15 are probably the least
likely to harbor IMBHs.  Rather, dynamical heating by even a modest
IMBH is likely to lead to a cluster containing a fairly extended core.
Comparing the outward energy flux from stars relaxing inward in the
Bahcall--Wolf (1976) cusp surrounding the IMBH (of mass $M_{BH}$) to
the outward flux implied by two-body relaxation at the cluster
half-mass radius, Heggie \etal estimate the equilibrium ratio of the
half-mass ($R_h$) to the core ($R_c$) radius.  Calibrating to
simulations, they conclude that
\begin{equation}
	\frac{R_h}{R_c} ~\sim~ 0.23 \left(\frac{M}{M_{BH}}\right)^{3/4}.
\label{Eq:rcrh}
\end{equation}
Trenti (2008) has suggested that the imprint of this process can be
seen in his ``isolated and relaxed'' sample of Galactic clusters
having relaxation times less than 1 Gyr, a half-mass to tidal radius
ratio $R_h/R_t < 0.1$, and an orbital ellipticity of less than 0.1.
Roughly half of the clusters in this sample have core radii
substantially larger than would be expected on the basis of simple
stellar dynamics and binary heating.  However, Hurley (2007) has
argued out that such anomalously large core to half-mass ratios may
also be explained by the presence of a stellar-mass BH binaries
heating the cores of these clusters (see also Mackey \etal 2007).


\section{Formation of Intermediate-mass Black Holes}\label{sec:formation}

Accepting without further debate the still sketchy observational
evidence for IMBHs in globular clusters, we now turn to the question
of how such black holes might have formed.  The leading possibilities
are (1) they are primordial, the result of stellar evolution in
supermassive population III stars, (2) they are the result of runaway
stellar collisions in young dense star clusters, and (3) they are the
result of mergers of BHs over the lifetime of a cluster.  We focus
here on the latter two possibilities, since the formation mechanisms
involved are arguably much more relevant to the physics of dense
stellar environments.

The likelihood of multiple stellar collisions in dense stellar systems
was first demonstrated in $N$-body simulations of dense clusters
(Portegies Zwart \etal 1999, 2004; Portegies Zwart and McMillan
2002; McMillan and Portegies Zwart 2004), and later also in
Monte-Carlo models (Freitag \etal 2006).  Hydrodynamic simulations
indicate that, when massive stars collide in clusters, they are very
likely to merge.  The question then becomes ``What next?''
Unfortunately, while the dynamical processes leading to collision
runaways are simple and well known, there are several prominent
``missing links'' in the chain of reasoning starting from a young
stellar system and ending with a massive black hole, all involving key
aspects of the physics of massive stars.


\subsection{Mass Segregation and Runaway Mergers}\label{sec:runaway}

The dynamics of runaway mergers is straightforward.  Massive stars
sink to the cluster core on a time scale $t_s$ (Eq.~\ref{eq:ts}).  The
result, for a realistic (Kroupa 2001) stellar mass spectrum, is the
formation of a dense sub-core of massive stars in a time
$t_{cc}\sim0.2t_r$.  Here, $t_r$ can be the half-mass relaxation time
for a small system, or the core relaxation time for a larger one, as
discussed below.  High densities in the core lead to collisions and
mergers, which naturally involve the most massive stars, and the
process runs away, with one collision product growing rapidly in mass
and radius and outstripping the competition.  The resultant merger
(and, we assume, black hole) mass is in the IMBH range.  For a runaway
to occur, the collision process must complete before the first
supernovae occur, that is, within 3--5 Myr.  This requires short
relaxation times, or, equivalently, high cluster densities.

\begin{figure}
  \vskip 0.2in
  \centerline{\psfig{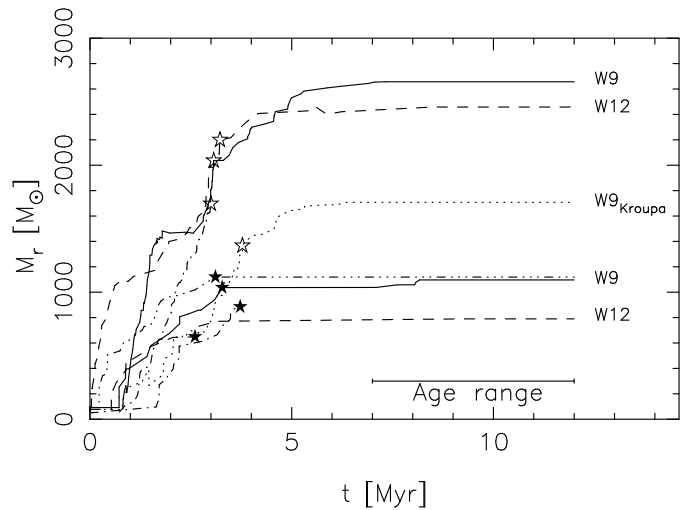}}
  \caption{Growth in mass of the runaway star for simulations
           performed with {\tt Starlab} (Portegies Zwart \etal 2001)
           and {\tt NBODY4} (Aarseth 1999, Baumgardt and Makino 2003).
           The choice of initial concentration is labeled by the
           dimensionless central potential, where W12(9) implies a
           King parameter $W_0 = 12(9)$.  The upper curves are for
           {\tt NBODY4}; the lower ones for {\tt Starlab}.  The star
           symbols indicate the moment when a supernova occurs,
           typically around 3\,Myr.}
  \label{fig:runaway}
\end{figure}

Figure \ref{fig:runaway} shows a typical set of results, obtained by
Portegies Zwart \etal (2004) in their $N$-body study of the M82
clusters discussed earlier.  The different runs represent a broad
range of initial conditions, with and without initial binaries and
with both Salpeter (1955) and Kroupa (2001) mass functions, for
systems of 128k--585k stars.  The runaway masses in {\tt NBODY4} are
generally larger than those in {\tt Starlab}, since {\tt NBODY4}
adopts systematically larger stellar radii and hence collision cross
sections, for the most massive stars.  Comparable results (in both
time scale and in total runaway mass) have been obtained in
Monte-Carlo simulations (e.g. Freitag \etal 2006).

According to the dynamical models, a runaway always occurs if a
central core of massive stars ($m \apgt 20 \msun$) can form by mass
segregation before the first supernova occurs.  In small systems,
having $t_{rh} \aplt 25$ Myr, essentially all the massive stars can
reach the center in the time available.  The merger mass fraction, for
a typical (Kroupa 2001) initial mass function, is $\sim 0.01$
(Portegies Zwart and McMillan 2002).  In large systems, only a
fraction of massive stars can reach the center before exploding as
supernovae.  Freitag \etal (2006) report a somewhat smaller mass
fraction for the IMBH in their Monte-Carlo models.  McMillan and
Portegies Zwart (2004) find that, for $t_{rh} \apgt 25$, the merger
fraction is expected to scale as $t_{rh}^{-1/2}$.


\subsection{Evolution of the Merger Product}\label{sec:evolution}

Freitag \etal (2006) find that direct mass loss due to the collision
itself is generally unimportant, amounting to less than $\sim10$\% of
the total mass in most cases.  However, the problem of determining the
evolution of a possibly rapidly rotating collision product, from its
initial non-equilibrium state back to the anomalous main sequence and
beyond, presents significant computational challenges (e.g. Sills
\etal 2003).

Perhaps the greatest uncertainty in the runaway process has to do with
mass loss from supermassive stars.  Massive main sequence stars are
well known to have very strong winds, and these compete with
collisions in setting the mass of the final object (e.g. Vanbeveren
\etal 2007; Belkus \etal 2007).  Portegies Zwart \etal (1999)
recognized that the specifics of the mass loss prescription completely
control the outcome of the collision runaway.  Applying most of the
mass loss late, at the terminal-age main sequence, permits the process
to build massive stars, and this prescription has (unfortunately) been
adopted in most dynamical models to date.  A more realistic treatment
of mass loss (based on Langer \etal 1994) can significantly reduce the
mass of the final collision product, while (probably unrealistic)
early mass loss can shut the process down completely.

By way of calibration, we note that dynamical models typically predict
the accretion of $\sim10^3\msun$ of material in $\sim10^6$ yr, for a
net accretion rate of $\sim10^{-3}\msun$/yr.  This is comparable to
the mass loss rates reported for several massive stars, but
substantially less than the largest rates known, e.g. the
$\sim0.1\msun$/yr inferred during the decade around the ``great
outburst'' of Eta Carinae in 1843 (Morris \etal 1999).  Belkus \etal
(2007) find that massive (300--1000 \msun) solar-metallicity stars end
their lives as relatively low-mass (40--50 \msun) black holes, but
comparably massive stars in low-metallicity clusters may give rise to
IMBHs in the 150--200 {\msun} range.  Yungelson \etal (2008) find a
final black hole mass of $\sim150\msun$ for a star with initial mass
$1000\msun$.

These studies suggest that it may be difficult to retain enough mass
to form an IMBH, although recently, Suzuki \etal (2007) have performed
simulations of collisionally merged stars, using a more realistic
mass-loss model than in most earlier dynamical studies, and find that,
because of the extended envelopes of the merged systems, most
collisions are expected to occur early, during the Kelvin-Helmholtz
contraction phase, before mass loss by stellar winds become
significant. They conclude that stellar mass loss does not prevent the
formation of massive stars with masses up to $\sim1000\msun$.
However, this in turn contrasts sharply with the work of Glebbeek
(2008; see also Gaburov \etal 2008, Glebbeek \etal 2008), who follows
the evolution of a series of collision products drawn from $N$-body
simulations, computing in detail the return to the main sequence and
the subsequent evolution of each merged star.  He concludes that, at
least in stars of solar or near-solar metallicity, strong line-driven
winds will expel much of the accumulated envelope, resulting in a
Wolf-Rayet star of mass considerably less than 100 {\msun} by the time
a supernova occurs.  Obviously, the final word on this important but
poorly understood process has yet to be written.

Finally, even if mass loss were not a factor, we remind the reader
that the assumption that a $1000\msun$ star will ultimately form a
$1000\msun$ black hole is also largely a matter of conjecture (see
Heger 2008).


\subsection{Connection with Globular Clusters}

Consider a compact, centrally concentrated young star cluster, with a
central density high enough for the runaway collision scenario to
operate.  For a $\sim10^6\msun$ system, the constraint on the
relaxation time presented in \S\ref{sec:runaway} requires a mean
density of $\sim5\times10^7\msun\,{\rm pc}^{-3}$, for a half-mass
radius of $\sim 0.2$ pc.  One might reasonably wonder what such a
highly concentrated, dense system has to do with the relatively
low-density globular clusters seen today.  Can runaway collisional
processes shortly after formation account for the IMBHs that may exist
in the Galactic globular cluster system?  Based on simulations of many
aspects of the problem, we can construct a plausible scenario in which
collisionally formed IMBHs might now reside in globular clusters.

After the IMBH has formed, numerous evolutionary processes combine
both to expand the cluster half-mass radius $R_h$ and also to reduce
its central concentration, as measured by the ratio $R_h/R_c$
(Eq.~\ref{Eq:rcrh}).  If the cluster is not mass segregated at birth,
the dynamical effects of segregation by massive stars and their
remnants will cause the concentration to decrease significantly, and
will also result in a modest overall expansion of the cluster (Merritt
\etal 2004; Mackey \etal 2007).  At the same time, mass loss due to
stellar evolution drives further overall expansion, by a factor of
2--3 (Takahashi and Portegies Zwart 2000; Mackey \etal 2007).
Vesperini \etal (2008) find that mass loss from the cores of initially
mass-segregated clusters is particularly effective in reducing the
central concentration.  Finally, heating due to the central IMBH
(\S\ref{sec:indirect}) again reduces the concentration and causes
significant expansion of $R_h$ over the lifetime of the cluster; the
half-mass radii of the model clusters considered by Baumgardt \etal
(2005) expanded by factors of 5--7.

Taken together, these figures suggest that the collision runaway
scenario could indeed account for an IMBH in a system the size of a
typical globular cluster---if one is ever confirmed!


\subsection{Systems of Black Holes}

We now briefly consider the opposite extreme, in which the relaxation
time is long enough that the massive stars form stellar mass black
holes ``in place,'' before significant dynamical evolution can occur.

The evolution of black-hole systems in low-density clusters has been
studied by a number of authors.  Kulkarni \etal (1993) and Sigurdsson
and Hernquist (1993) considered the dynamics of a population of black
holes in an idealized star cluster.  They found that mass segregation
rapidly transports the black holes to the cluster core, where the
black hole subsystem evolves rapidly, forming binaries that ultimately
eject most of the black holes from the cluster.  Both papers concluded
that few observable stellar-mass black holes would be expected in the
Galactic globular cluster system today.  Subsequently, Portegies Zwart
and McMillan (2000) performed $N$-body simulations, and generally
verified these findings, but also concluded that the ejected
black-hole binaries could be significant sources of gravitational
waves (GW) in the LIGO band.

These studies were idealized in several important ways.  They were not
dynamically self-consistent (the $N$-body simulations combined many
small-N simulations to achieve better statistics), they did not
include a spectrum of black hole masses, and they did not include
post-Newtonian relativistic effects (all used the Peters and Mathews
[1963] formula for GW energy losses).  Miller and Hamilton (2002)
pointed out that a sufficiently massive black hole, at the top end of
the stellar black hole mass spectrum, or perhaps the result of a
``failed'' runaway merger, could survive ejection by dynamical
interactions, and proposed a mechanism whereby black holes in binary
systems would merge by the emission of GW to form successively more
massive objects, ultimately resulting in an IMBH.  However, it now
seems unlikely that this process can operate efficiently in the face
of the large GW recoil velocities found in recent numerical
simulations of black-hole mergers (see Centrella 2008 for a review).

O'Leary \etal (2006) carried out Monte-Carlo simulations of a
population of black holes (roughly 50\% binaries) drawn from a
population synthesis calculation starting from realistic distributions
of stellar and binary properties.  They included approximate
post-Newtonian effects in their model, and found that, although modest
IMBHs ($\sim100\msun$, up to a maximum of $\sim600\msun$) did form, in
most cases the recoil speed substantially exceeded the escape speed
from the cluster, and that the most likely outcome was evaporation of
the entire black hole subsystem.  More recent studies, incorporating
semi-analytic treatments of GW recoil calibrated to numerical
relativity calculations, greatly strengthen this latter result
(Amaro-Seoane \etal 2008).  The recoil speeds in these simulations are
substantially higher than are obtained in the post-Newtonian (up to
3.5PN) approximation, with the result that virtually none of the BH
merger products are expected to remain bound to the cluster.  This
finding effectively effectively closing the BH merger avenue for IMBH
formation and growth, unless an already very massive ($>100\msun$)
seed BH can form by some other means.

Although this scenario does not appear to offer a viable channel for
IMBH production, the many BH binary mergers present potentially rich
targets for the enhanced and advanced LIGO detectors.  O'Leary \etal
(2007) infer an advanced LIGO detection rate of up to hundreds of
events per year, broadly consistent with the range given previously by
Portegies Zwart and McMillan (2000).  However, these estimates are
plagued by large uncertainties, due principally to the spread in
structural properties of the parent clusters, amplified by the scaling
of the GW emission process.  Specifically, using the distributions of
BH binary energy and eccentricity reported by Portegies Zwart and
McMillan (2000) and the Peters and Matthews (1963) radiation formula,
it is readily shown that the characteristic time scale at which the
distribution of GW merger times peaks is proportional to $m_{bh}^5
(M/R)^{-4}$, where $m_{bh}$ is a typical stellar BH mass and $M$ and
$R$ are, respectively, the mass and radius of the cluster.  The
uncertainty is compounded when these numbers are scaled up from
individual cluster simulations to integrated populations of clusters
across the universe.  Nevertheless, the very fact that the current
best estimates predict many detections by upcoming ground-based
instruments means that, whether or not the predicted events are
actually seen, the observations should place strong constraints on the
cluster formation history of the universe.

Recently, Mackey \etal (2007) have reported a set of fully
self-consistent (Newtonian) dynamical simulations of young dense
clusters.  They find that the combination of stellar mass loss and
dynamical heating has a significant dynamical impact on the cluster,
causing the core (and, to a lesser extent, the cluster as a whole) to
expand in a manner strikingly similar to the core radius--age relation
observed in the young clusters in the LMC (Mackey and Gilmore 2003).
However, because of the expansion, the black hole interaction rate
drops sharply at late times, and a substantial population of black
holes remains after 10 Gyr.  If, as seems plausible, this general
dynamical result scales to globular-cluster-sized systems, it suggests
that the absence of evidence of stellar-mass black holes is not
evidence of absence, but simply reflects their low current interaction
rate with other cluster members.


\section{Discussion}

There is growing, but arguably still inconclusive, evidence that some
globular star clusters may harbor massive black holes in their cores.
Perhaps the strongest cases come from dynamical studies of G1 in M31
and $\omega$ Centauri in the Milky Way, but even here the conclusions
are tainted to some degree by the suspicion that these massive
clusters may actually be the stripped cores of dwarf spheroidal
galaxies, in which case they shed little light on the physics of young
star clusters or the collisional processes that may give rise to IMBHs
in stellar systems.  Indirect evidence based on cluster structural
parameters and central density and velocity profiles is almost as
compelling, and follow-up studies of the central velocity structure of
``real'' globular clusters have the potential to revolutionize the
debate on this subject.

From a dynamical modeler's perspective, at least, the possibility that
collisions in sufficiently dense young clusters might lead to runaway
mergers offers the most interesting path to IMBHs in globular
clusters.  Extensive dynamical simulations leave no doubt that stellar
collisions occur in sufficiently dense systems, and simple arguments
lead to the inevitable conclusion that there is easily enough mass
potentially available in these systems to produce objects with masses
exceeding $1000\msun$.  In many ways, this process provides a
``natural'' mechanism for IMBH formation, but critical aspects of the
evolution of the merger product---specifically, the role of stellar
winds and the end result of the evolution of very massive
stars---remain poorly understood.  At present, it appears that stars
with metallicities comparable to that of the Sun may be unable to
retain enough mass to form an IMBH, but that collisions in
low-metallicity systems at early times may conceivably have led to
runaways in the progenitors of today's globular clusters.

The role of dense stellar systems as potential gravitational wave (GW)
source factories has been recognized for many years, and dynamical
interactions in these systems may produce a wide variety of
interesting GW sources.  We confine ourselves here to systems
containing IMBHs formed by dynamical means, emphasizing again that all
of the following estimates should be treated with caution (if not
outright skepticism), combining as they do the major uncertainties
inherent in both the formation mechanism of the IMBHs themselves and
the spread in properties of their parent clusters.

\begin{itemize}

\item{\em BH--IMBH inspirals}~~Intermediate-mass-ratio inspirals
  (IMRIs) involving IMBHs formed at the centers of dense star clusters
  may be important sources of low-frequency GWs for LISA (Barack and
  Cutler 2004, Gair \etal 2004, Will 2004; Amaro-Seoane \etal 2007).
  These could plausibly arise through a variety of mechanisms in star
  clusters, including hardening of an IMBH binary, exchange
  interactions between a BH binary and an IMBH, tidal capture of the
  progenitor star, three-body resonances, or even direct capture via
  the emission of GW during a close encounter (see Mandel \etal 2007
  for a review).  With the best current estimates for merger rates and
  signal processing performance, LISA could detect $\sim 10$ such IMBH
  IMRIs during a three-year run (Miller 2002, Mandel \etal 2007).
  Advanced LIGO may also be able to detect IMRIs of stellar-mass
  compact objects into low-mass ($\aplt 400 \msun$) IMBHs in globular
  clusters, at a rate of $\sim10$ events per year (Brown \etal 2007,
  Mandel \etal 2007).

  ~~~~The weakness of the gravitational radiation from a typical BH
  inspiral into an IMBH means that the signal from tens of thousands
  of orbits must be integrated in order to to detect the source.
  Therefore, knowledge of the waveform, and especially the probable
  distribution of orbital eccentricities as determined by simulations,
  is essential to their detection.  Mandel \etal (2007) conclude that
  IMRIs in clusters are unlikely to exhibit significant eccentricity
  by the time they enter the the LISA band.  However, even for
  low-eccentricity systems, the theoretical IMRI waveforms are not
  well known, and this uncertainty may still lead to difficulties in
  determining the parameters of observed systems.

\item{\em IMBH--IMBH binaries}~~There are also hints that in some
  circumstances more than one IMBH might form in such a cluster
  (G\"urkan \etal 2006).  Two IMBHs in a cluster core are expected to
  form a binary and merge rapidly by GW emission, in as little as a
  few million years.  Such IMBH--IMBH mergers would be powerful and
  unique probes of the star formation history of the universe.
  Fregeau \etal (2006) estimate that LISA could detect tens of
  IMBH--IMBH inspirals per year, while advanced LIGO might observe 10
  merger and ringdown events per year.  However, they find that most
  such mergers occur at redshift $z\sim1$, presumably from relatively
  high-metallicity progenitors, which may provide the least favorable
  environment for IMBH production (see \S\ref{sec:evolution}).

\item{\em IMBH--SMBH mergers}~~On larger scales, IMBHs in clusters
  close enough to the center of a galaxy may spiral in and merge with
  the central supermassive black hole (Miller 2005, Matsubayashi \etal
  2007).  Taking the simple dynamical estimates presented in
  Sec.~\ref{sec:runaway} at face value and integrating over a cluster
  population representative of the inner Galactic bulge, Portegies
  Zwart \etal (2006) infer a IMBH--SMBH merger rate in the Galactic
  center of a few per 100 Myr.  The scaling of these rates to larger
  galactic nuclei, and hence the net merger rate potentially
  observable by LISA, is not well known, although Matsubayashi \etal
  (2007) estimate a net LISA event rate of $\sim10$ per year.

\end{itemize}


\section*{Acknowledgements}

This work has been suppported by NASA grants NNG04GL50G and
NNX07AG95G, and by NSF grant AST-0708299.


\section*{References}

\begin{harvard}

\item[] Aarseth S J 1999 {\em PASP} {\bf 111} 1333

\item[] Aarseth S J 2003 {\em The Gravitational $N$-Body Problem}
  (Cambridge: Cambridge University Press)

\item[] Amaro-Seoane P, Gair J R, Freitag M, Miller M C, Mandel I,
  Cutler C J and Babak S 2007 {\em Classical and Quantum Gravity} {\bf
    24} 113

\item[] Amaro-Seoane P, Kupi G and Fregeau J M 2008 in preparation

\item[] Antonov V A 1962 {\em Vestn Leningr Gros Univ} {\bf 7} 135

\item[] Bahcall J N and Wolf R A 1976 {\em ApJ} {\bf 209} 214

\item[] Barack L and Cutler C 2004 {\em Phys Rev D} {\bf 70} 12

\item[] Baumgardt H and Makino J 2003 {\em MNRAS} {\bf 340} 227

\item[] Baumgardt H, Hut P, Makino J, McMillan S L W, and Portegies
  Zwart S F 2003a {\em ApJ} {\bf 582} L21

\item[] Baumgardt H, Makino J, Hut P, McMillan S L W, and Portegies
  Zwart S F 2003b {\em ApJ} {\bf 589} 25

\item[] Baumgardt H, Makino J and Ebisuzaki T 2004 {\em ApJ} {\bf 613}
  1133

\item[] Baumgardt H, Makino J and Hut P 2005 {\em ApJ} {\bf 620} 238

\item[] Belkus H, Van Bever J and Vanbeveren D 2007 {\em ApJ} {\bf
  659} 1576

\item[] Belleman R G, Bédorf J and Portegies Zwart S F 2008 {\em New
    Astronomy} {\bf 13} 103

\item[] Brown D A, Brink J, Fang H, Gair J R, Li C, Lovelace G, Mandel
  I and Thorne K S 2007 {\rm PRL} {\bf 99} 201102

\item[] Centrella J 2008 in STScI Spring Symposium ``Black Holes'' eds M
  Livio and A M Koekemoer (Cambridge: Cambridge University Press)

\item[] Church R 2006 PhD thesis, Cambridge University

\item[] Cohn H N 1980 {\em ApJ} {\bf 242} 765

\item[] Dull J D, Cohn H N, Lugger P M, Murphy B W, Seitzer P O,
  Callanan P, J Rutten R G M and Charles P A 1997 {\em ApJ} {\bf 481} 267

\item[] Ebisuzaki T, Makino J, Taiji M, Fukishige T, Sugimoto D, Ito
  T and Okumura S K 1993 {\em PASJ} {\bf 45} 269

\item[] Eggleton P P, Fitchett M J and Tout C A 1989 {\em ApJ} {\bf 347} 998

\item[] Eggleton P P 2006 {\em Evolutionary Processes in Binary and
  Multiple Stars} (Cambridge: Cambridge University Press)

\item[] Freeman K C 1993 in {\em Galactic Bulges} IAU Symposium 153 ed
  H Dejonghe and H J Habing (Dordrecht: Kluwer) p 263

\item[] Fregeau J M, G\"urkan M A, Joshi K J and Rasio F A 2003 {\em ApJ}
  {\bf 593} 772

\item[] Fregeau J M, Larson S L, Miller M C, O'Shaughnessy R and Rasio
  F A 2006 {\em ApJ} {\bf 646} L135

\item[] Gair J R, Barack L, Cutler C, Creighton T, Larson S L, Phinney
  E S and Vallisneri M 2004 {\em Class Quant Grav} {\bf 21} S1595

\item[] Giersz M 1998 {\em MNRAS} {\bf 298} 1239

\item[] Giersz M and Heggie D C 2008 arXiv:astro-ph/0801.3968

\item[] G\"urkan M A, Fregeau J M and Rasio F A 2006 {\em ApJ} {\bf 640} 39

\item[] Freitag M and Benz W 2005 {\em MNRAS} {\bf 358} 1133

\item[] Freitag M, Rasio F A and Baumgardt H 2006 {\em MNRAS} {\bf 368} 121

\item[] Freitag M, G\"urkan M A and Rasio F A 2006 {\em MNRAS} {\bf 368} 141

\item[] Gaburov E, Gualandris A and Portegies Zwart S F 2008 {\em MNRAS}
  {\bf 384} 376

\item[] Gebhardt K, Rich R M R and Ho L C 2002 {\em ApJ} {\bf 578} L41

\item[] Gebhardt K, Rich R M R and Ho L C 2005 {\em ApJ} {\bf 634} 1093

\item[] Gerssen J, van der Marel R P, Gebhardt K, Guhathakurta P,
  Peterson R and Pryor C 2002 {\em AJ} {\bf 124} 327

\item[] Gerssen J, van der Marel R P, Gebhardt K, Guhathakurta P,
  Peterson R and Pryor C 2003 {\em AJ} {\bf 125} 376

\item[] Glebbeek E 2008 {\em Ph.D Thesis, Utrecht University}

\item[] Glebbeek E, Gaburov E, de Mink S E, Pols O and Portegies Zwart
  S F 2008 in preparation

\item[] Harris W E 1996 {\em AJ} {\bf 112} 1487 (2003 revision of the catalog)

\item[] Heger A 2008 in STScI Spring Symposium ``Black Holes'' eds M
  Livio and A M Koekemoer (Cambridge: Cambridge University Press)

\item[] Heggie D C 1975 {\em MNRAS} {\bf 173} 729

\item[] Heggie D C, Hut P, Minishige S , Makino J and Baumgardt H 2007
  {\em PASJ} {\bf 59} 507

\item[] Hurley J R, Pols O R and Tout C A 2000 {\em MNRAS} {\bf 315}
  543

\item [] Hurley J R 2007 {\em MNRAS} {\bf 379} 93

\item[] Inagaki S 1985 in IAU Symposium 113 {\em Dynamics of Star
  Clusters} eds J Goodman and P Hut (Dordrecht: Reidel)

\item[] Joshi K J, Rasio F A and Portegies Zwart S F 2000 {\em ApJ}
  {\bf 540} 969

\item[] Kaaret P, Prestwich A H, Zezas A, Murray S S, Kim D-W, Kilgard
  R E, Schlegel E M and Ward M J 2001 {\em MNRAS} {\bf 321} L29

\item[] Kroupa P 2001 {\em MNRAS} {\bf 322} 231

\item[] Kulkarni S R, Hut P and McMillan S L W 1993 {\em Nature} {\bf
  364} 421

\item[] Langer N, Hamann W-R, Lennon M, Najarro F, Pauldrach A W A and
  Puls J 1994 {\em Astronomy and Astrophysics} {\bf 290} 819

\item[] Lombardi J C 2007 MODEST-8 workshop, Bonn, December 2007

\item[] Maccarone T J, Kundu A, Zepf S E and Rhode K L 2007 {\em
  Nature} {\bf 445} 183

\item[] Mackey A D and Gilmore G F 2003 {\em MNRAS} {\bf 338} 120

\item[] Mackey A D, Wilkinson M I, Davies M B and Gilmore G F 2007
  {\em MNRAS} {\bf 379} 40

\item[] Makino J and Aarseth S J 1992 {\em PASJ} {\bf 44} 141

\item[] Makino J, Taiji M, Ebisuzaki T and Sugimoto D 1997 {\em ApJ}
  {\bf 480} 432

\item[] Mandel I, Brown D A, Gair J R and Miller M C 2007
  arXiv:astro-ph/0705.0285

\item[] Matsubayashi T, Makino J and Ebisuzaki T 2007 {\em ApJ} {\bf
  656} 879

\item[] Matsumoto H Tsuru T G 1999 {\em PASJ} {\bf 51} 321

\item[] Matsumoto H, Tsuru T G, Koyama K, Awaki H, Canizares C R,
  Kawai N, Matsushita S and Kawabe R 2001 {\em ApJ} {\bf 547} L25

\item[] McCrady N, Gilbert A M and Graham J R 2003 {\em ApJ} {\bf 596}
  240

\item[] McMillan S L W, Hut P and Makino J 1991 {\em ApJ} {\bf 372}
  111

\item[] McMillan S L W and Hut P 1994 {\em ApJ} {\bf 427} 793

\item[] McMillan S L W and Portegies Zwart S F 2004 in {\em Massive
  Stars in Interactive Binaries} Astronomical Society of the Pacific
  Conference Series 367 ed N St-Louis and A F J Moffat (San Francisco:
  Astronomical Society of the Pacific 2007) p 697

\item[] Merloni A, Heinz S and di Matteo T 2003 {\em MNRAS} {\bf 345}
  1057

\item[] Merrit D and Ferrarese L 2001 {\em ApJ} {\bf 547} 140

\item[] Merrit D, Piatek S, Portegies Zwart S F and Hemsendorf M 2004
  {\em ApJ} {\bf 608} 25

\item[] Meylan G, Sarajedini A, Jablonka P, Djorgovski S G, Bridges T
  and Rich R M 2001 {\em AJ} {\bf 122} 830

\item[] Miller M C 2002 {\em ApJ} {\bf 581} 438

\item[] Miller M C and Hamilton D P 2002 {\em MNRAS} {\bf 330} 232

\item[] Miller M C and Colbert E J M 2004 {\em Int J Mod Phys D} {\bf
  13} 1

\item[] Miller M C 2005 {\em ApJ} {\bf 618} 426

\item[] Morris P W, Waters L B F M, Barlow M J, Lim T, de Koter A,
  Voors R H M, Cox P, de Graauw Th, Henning Th, Hony S, Lamers H J G L
  M, Mutschke H and Trams N R 1999 {\em Nature} {\bf 402} 502

\item[] Noyola E and Gebhardt K 2006 {\em AJ} {\bf 132} 447

\item[] Noyola E, Gebhardt K J, and Bergmann M 2006 in {\em New
  Horizons in Astronomy: Frank N Bash Symposium} ASP Conference Series
  Vol 352 ed S J Kannappan S Redfield J E Kessler-Silacci M Landriau
  and N Drory (San Francisco: Astronomical Society of the Pacific) p
  269

\item[] Noyola E, Gebhardt K and Kissler-Patig M 2008 in preparation

\item[] Noyola E 2008 in STScI Spring Symposium ``Black Holes'' eds M
  Livio and A M Koekemoer (Cambridge: Cambridge University Press)

\item[] O'Leary R M, Rasio F A, Fregeau J M, Ivanova N and
  O'Shaughnessy R 2006 {\em ApJ} {\bf 637} 937

\item[] Peters P C and Mathews J 1963 {\em Phys Rev D} {\bf 131} 345

\item[] Phinney E S 1993 in {\em Structure and Dynamics of Globular
  Clusters} ed S Djorgovski and G Meylan (San Francisco: Astronomical
  Society of the Pacific) p 141

\item[] Pooley D and Rappaport S 2006 {\em ApJ} {\bf 644} 45

\item[] Portegies Zwart S F, Makino J, McMillan S L W and Hut P 1999
  {\em Astronomy and Astrophysics} {\bf 348} 117

\item[] Portegies Zwart S F and McMillan S L W 2000 {\em ApJ} {\bf
  528} 17

\item[] Portegies Zwart S F, McMillan S L W, Hut P and Makino J 2001
  {\em MNRAS} {\bf 321} 199

\item[] Portegies Zwart S F and McMillan S L W 2002 {\em ApJ} {\bf 576}
  899

\item[] Portegies Zwart S F, Bellmann R G and Geldof P M 2007 {\em New
  Astronomy} {\bf 12} 641

\item[] Portegies Zwart S F, Baumgardt H, Hut P, Makino J and McMillan
  S L W 2004 {\em Nature} {\bf 428} 724

\item[] Portegies Zwart S F, Baumgardt H, McMillan S L W, Makino J,
  Hut P and Ebisuzaki T 2006 {\em ApJ} {\bf 641} 319

\item[] Portegies Zwart S F, McMillan S L W, O'Nuall\'ain B, Heggie D
  C, Lombardi J, Hut P, Banerjee S, Belkus H, Fragos T, Fregeau J,
  Fujii M, Gaburov E, Harfst S, Izzard R, Juri\'c M, Justham S, Teuben
  P, van Bever J, Yaron O and Zemp M 2008 to appear in {\em Simulation
    of Multiphysics Multiscale Systems} (International Conference on
  Computational Science, ed R Belleman)

\item[] Salpeter E E 1955 {\em ApJ} {\bf 121} 161

\item[] Sigurdsson S and Hernquist 1993 {\em Nature} {\bf 364} 423

\item[] Sills A, Deiters S, Eggleton P P, Freitag M, Giersz M, Heggie
  D, Hurley J, Hut P, Ivanova N, Klessen R S, Kroupa P, Lombardi J C,
  McMillan S L W, Portegies Zwart S F and Zinnecker H 2003 {\em New
  Astronomy} {\bf 8} 605

\item[] Soria R 2006 in {\em Populations of High Energy Sources in
  Galaxies} IAU Symposium 230 ed E J A Meurs and G Fabbiano
  (Cambridge: Cambridge University Press) p 473

\item[] Soria R 2007 {\em Astrophysics and Space Science} {\bf 311} 213

\item[] Spitzer L Jr 1969 {\em ApJL} {\bf 158} L139

\item[] Spitzer L Jr 1987 {\em Dynamical Evolution of Globular
  Clusters} Princeton University Press

\item[] Strohmayer T E and Mushotsky R F 2003 {\em ApJ} {\bf 586} L61

\item[] Suzuki T K, Nakasato N, Baumgardt H, Ibukiyama A, Makino J and
  Ebisuzaki T 2007 {\em ApJ} {\bf 668} 435
 
\item[] Takahashi K and Portegies Zwart S F 2000 {\em ApJ} {\bf 535}
  759

\item[] Tremaine S, Gebhardt K, Bender R, Bower G, Dressler A, Faber S
  M, Fillipenko A V, Green R, Grillmair C, Ho L C, Kormendy J, Lauer
  T, Magorrian J, Pinkney J and Richstone D 2002 {\em ApJ} {\bf 574}
  740

\item[] Trenti M 2008 in STScI Spring Symposium ``Black Holes'' eds M
  Livio and A M Koekemoer (Cambridge: Cambridge University Press)

\item[] Ulvestad J S, Greene J and Ho L C 2007 {\em ApJ} {\bf 661} 151

\item[] Vanbeveren D, Belkus H, Van Bever J and Mennekens N 2007
  arXiv:astro-ph/07123343

\item[] Vesperini E, McMillan S L W and Portegies Zwart S F 2008 in
  preparation

\item[] Will C M 2004 {\em ApJ} {\bf 611} 1080

\item[] Yungelson L R, van den Heuvel E P J, Vink J S, Portegies Zwart
  S F and de Koter A 2008 {\em Astronomy and Astrophysics} {\bf 477}
  223

\end{harvard}

\end{document}